\documentclass[10pt,aps,prmaterials,twocolumn,longbibliography,nobibnotes,showkeys]{revtex4-2}

\usepackage{graphicx}
\usepackage{dcolumn}
\usepackage{bm}
\usepackage{color}
\usepackage{booktabs}
\usepackage{amssymb}
\usepackage{amsmath}

\begin{document}
\title{First-principles characterization of native defects and oxygen impurities in GaAs}
\author{Khang Hoang}
\email{khang.hoang@ndsu.edu}
\affiliation{Center for Computationally Assisted Science and Technology \& Department of Physics, North Dakota State University, Fargo, North Dakota 58108, United States}

\date{\today}

\begin{abstract}

We present a systematic investigation of native point defects and oxygen impurities in GaAs using hybrid functional calculations. Defects are characterized by their structural, electronic, and optical properties. Under thermodynamic equilibrium, dominant native defects are Ga antisites (Ga$_{\rm As}$), As antisites (As$_{\rm Ga}$), and/or Ga vacancies ($V_{\rm Ga}$) in which As$_{\rm Ga}$ and $V_{\rm Ga}$ are charge-compensating defects under As-rich conditions. On the basis of the defect transition levels, the isolated As$_{\rm Ga}$ can be identified with the $EL2$ center reported in experiments. The defect, however, has a negligible nonradiative electron capture cross section and thus cannot be the ``main electron trap'' as commonly believed. We find that GaAs can have multiple O-related defect centers, especially when prepared under As-rich conditions. The quasi-substitutional O impurity (O$_{\rm As}$) and its complex with two As$_{\rm Ga}$ defects (O$_{\rm As}$-2As$_{\rm Ga}$) both have a metastable and paramagnetic middle (neutral) charge state; however, only the latter can be identified with the experimentally observed Ga--O--Ga or ``OX'' center. These two defects have large nonradiative electron capture cross sections and can be effective carrier traps or recombination centers, which has important implications for materials design.

\end{abstract}

\pacs{}

\maketitle


\section{Introduction}\label{sec;intro}

GaAs-based materials have been used in electronics, optoelectronics, and photovoltaics~\cite{delAlamo2011Nature,Krishnamoorthy1996IEEE,Yablonovitch2012IEEE}. The ability to control and use defects has been key to the realization of such applications~\cite{Queisser1998Science}. Recently, defect complexes involving rare-earth and oxygen impurities in GaAs have also been of interest for quantum information science~\cite{Dietrich2016LPR,Higashi2020JAP,Fang2023OC,Fang2025JJAP}. Although defects in GaAs have been studied extensively experimentally and computationally over six decades~\cite{Blakemore1982JAP,Bourgoin1988JAP,Hurle1999JAP,Coutinho2020JPCM,Liu2025JAP}, an understanding remains incomplete. Unambiguous identification of defects in GaAs has been challenging as there are many possible defect centers; some of which may exist in stable and/or metastable configurations and some may share similar characteristics. Computational studies have played an important role in defect characterization. Yet, their predictive power largely depends on specific methods and procedures used in defect modeling. Using state-of-the-art first-principles defect calculations, this work revisits GaAs with a focus on native defects and oxygen impurities to develop a better understanding of defect physics in this technologically significant material.

Among many defect centers experimentally observed in GaAs, the so-called $EL2$ center has received much attention~\cite{Henry1977PRB,Mitonneau1979RPA,Bourgoin2001SST,Martin1977EL,Weber1982JAP,Lagowski1985APL,Omling1988PRB,Spicer1988JVST,Song1987APL,Bliss1992JAP,Liu1994APL,Luysberg1998JAP} due to, e.g., its role in growing semi-insulating GaAs crystals. This defect, often attributed to the isolated As antisite defect (As$_{\rm Ga}$), has been described as being abundant in As-excess GaAs materials, having energy levels at 0.50--0.54 eV and 0.75--0.77 eV above the valence-band maximum (VBM), and pinning the Fermi level near midgap~\cite{Henry1977PRB,Mitonneau1979RPA,Bourgoin2001SST,Martin1977EL,Weber1982JAP,Lagowski1985APL,Omling1988PRB,Spicer1988JVST,Song1987APL,Bliss1992JAP,Liu1994APL,Luysberg1998JAP}, among other characteristic properties~\cite{Dabrowski1989PRB}. It has often been referred to as the ``main electron trap'' in GaAs~\cite{Bourgoin2001SST,Martin1977EL,Mitonneau1979RPA,Weber1982JAP}. Ga vacancies ($V_{\rm Ga}$) have also been found as charge-compensating defects for As$_{\rm Ga}$~\cite{Bliss1992JAP,Liu1994APL,Luysberg1998JAP}. Other classic defect centers include $E1$, $E2$, $E3$, $E4$, and $E5$ observed in irradiated GaAs, with energy levels at 0.045, 0.14, 0.30, 0.76, and 0.96 eV, respectively, below the conduction-band minimum (CBM)~\cite{Pons1985JPC,Bourgoin1988JAP}. Taghizadeh et al.~\cite{Taghizadeh2022MSSP} found similar centers, denoted as $S1$ to $S5$, in sputtered GaAs at 0.046, 0.22, 0.30, 0.55, and 0.56 eV below the CBM, respectively; many of which are metastable. In addition to native point defects, oxygen impurities (either intentionally incorporated or unintentionally present) have also been well studied and known to often influence the electrical properties of semi-insulating GaAs, and multiple O-related defect centers been observed~\cite{Zhong1988APL,Schneider1989APL,Alt1989APLoxygen,Alt1990PRL,Song1990JAP,Skowronski1990APL,Neild1991APL,Skowronski1991APL,Skowronski1991JAP,Skowronski1992PRB,Jordan1992SST,Linde1995APL,Alt2007JAP}. One O-related center in particular (sometimes referred to as ``OX'' or ``O$_{\rm oc}$'') has been reported extensively and described in terms of a quasi-substitutional oxygen defect where the oxygen atom forms a Ga--O--Ga configuration and which exhibits the negative-$U$ character with a paramagnetic metastable middle charge state~\cite{Zhong1988APL,Schneider1989APL,Alt1989APLoxygen,Alt1990PRL,Song1990JAP,Skowronski1990APL,Skowronski1991APL,Neild1991APL,Skowronski1991JAP,Skowronski1992PRB,Jordan1992SST,Linde1995APL,Alt2007JAP}.  

Computationally, first-principles studies of native defects and oxygen impurities in GaAs have been widely reported~\cite{Dabrowski1988PRL,Chadi1988PRL,Dabrowski1989PRB,Zhang1990PRL,Mattila1996PRB,Poykko1996PRB,Taguchi1998PRB,Pesola1999PRB,Overhof2005PRB,Malouin2007PRB,Schultz2009MSMSE,Komsa2011PRB,Komsa2012JPCM,Colleoni2013ME,Chen2017PRB,Colleoni2013APL,Colleoni2016PRB,Wright2015PRB,Schultz2016PRB,Fluckey2026CMS}. They provided invaluable insights into the structural and electronic properties, including the bistability of certain defects. Examples of these previous contributions include Dabrowski and Scheffler~\cite{Dabrowski1988PRL} and Chadi and Chang~\cite{Chadi1988PRL} who identified the isolated antisite As$_{\rm Ga}$ as the $EL2$ center in its stable configuration, and found a metastable configuration $EL2^\ast$ in which the antisite As atom is displaced away from the Ga site.  Schultz and Anatole von Lilienfeld~\cite{Schultz2009MSMSE} assigned the $E1$ and $E2$ centers to the cation-anion divacancy ($V_{\rm Ga}$-$V_{\rm As}$) and the $E3$ center to the anion vacancy ($V_{\rm As}$). Pesola et al.~\cite{Pesola1999PRB} proposed O$_{\rm As}$-2As$_{\rm Ga}$, a complex of an O impurity and two As antisites, as a microscopic model for the Ga--O--Ga (``OX'') defect observed in experiments. 

Many of the previous studies were based on density-functional theory (DFT)~\cite{HK,KS} within the local-density (LDA) or generalized gradient approximation (GGA)~\cite{LDA1980,GGA}, whereas some others were based on a hybrid DFT/Hartree-Fock approach~\cite{Perdew1996JCP} or the $GW$ approximation~\cite{Hedin1965PR}. They all made use of supercell models with different sizes and of different atomic relaxation and finite-size correction schemes. Semilocal density functionals such as LDA/GGA are known to have limited predictive power~\cite{Freysoldt2014RMP}, however. In addition to the issue with the electronic structure (including the underestimation of the band gap), LDA/GGA-based calculations are known to often provide a poor description of defect structures and hence the energetics of even simple native point defects~\cite{Freysoldt2014RMP,Lyons2015,Hoang2021PRM}. We also note that, although structural and electronic properties of defects in GaAs have been widely studied, direct information on defect-related optical properties from calculations remains limited. Such information could provide additional means to computationally characterize the defects and help assess their possible role as carrier traps or recombination centers.

We herein report a study of native defects and oxygen impurities in GaAs using a hybrid DFT/Hartree-Fock approach~\cite{Perdew1996JCP} which has been proven to be able to provide a good description of defect physics in semiconductors~\cite{Freysoldt2014RMP,Hoang2018JPCM}. Large supercell models are employed to properly take into account local lattice relaxations and to reduce spurious defect--defect interactions. Optical properties of select defects are also investigated through the consideration of relevant defect-to-band radiative optical transitions and the calculation of nonradiative electron capture cross sections. Note that, while we systematically explore many native defects and oxygen impurities to produce a consistent dataset, we do not aim to search for all possible metastable configurations and defect complexes or to identify all defect centers observed in experiments, if that is ever possible. The results and discussion will focus mainly on As$_{\rm Ga}$ and O$_{\rm As}$-related defects where we demonstrate how, computationally, a combination of structural, electronic, and optical characterizations can lead to a more definitive identification of defects. 

\section{Methodology}\label{sec;method} 

Total-energy electronic structure calculations are based on DFT with the Heyd-Scuseria-Ernzerhof (HSE) functional~\cite{heyd:8207}, the projector augmented wave method~\cite{PAW2}, and a plane-wave basis set, as implemented in \textsc{vasp}~\cite{VASP2}. The Hartree-Fock mixing parameter is set to 0.28 and the screening length to the default value of 10 {\AA}. These parameters are chosen to reproduce the experimental band gap of GaAs. Defects in the GaAs host are simulated using 3$\times$3$\times$3 (216-atom) cubic supercells. In defect calculations, the lattice parameters are fixed to the calculated bulk values but all the internal coordinates are relaxed. The $\Gamma$ point is used for integration over the Brillouin zone. In all calculations, the plane-wave basis-set cutoff is set to 400 eV and spin polarization is included. All structural relaxations are performed with HSE and the force threshold is chosen to be 0.02 eV/{\AA}.   

A defect X in effective charge state $q$ is characterized by its formation energy, defined as~\cite{walle:3851,Freysoldt2014RMP}     
\begin{align}\label{eq:eform}
E^f({\mathrm{X}}^q)&=&E_{\mathrm{tot}}({\mathrm{X}}^q)-E_{\mathrm{tot}}({\mathrm{bulk}}) -\sum_{i}{n_i\mu_i} \\ %
\nonumber &&+~q(E_{\mathrm{v}}+\mu_{e})+ \Delta^q ,
\end{align}
where $E_{\mathrm{tot}}(\mathrm{X}^{q})$ and $E_{\mathrm{tot}}(\mathrm{bulk})$ are the total energies of the defect and bulk supercells; $n_{i}$ is the number of atoms of species $i$ that have been added ($n_{i}>0$) or removed ($n_{i}<0$) to form the defect; $\mu_{i}$ is the atomic chemical potential, representing the energy of the reservoir with which atoms are being exchanged. $\mu_{e}$ is the electronic chemical potential, i.e., the Fermi level, representing the energy of the electron reservoir, referenced to the VBM in the bulk ($E_{\mathrm{v}}$). Finally, $\Delta^q$ is the correction term to align the electrostatic potentials of the perfect bulk and defect supercells and to account for finite-size effects on the total energies of charged defects~\cite{Freysoldt,Freysoldt11}.

The chemical potentials of Ga and As are referenced to the total energy per atom of bulk Ga and As and vary over a range determined by the formation enthalpy of GaAs: $\mu_{\rm Ga} + \mu_{\rm As} = \Delta H ({\rm GaAs}) = -0.96$ eV (calculated at 0 K), comparable to the experimental value of $-0.74$ eV under ambient conditions~\cite{Wagman1982}. The extreme Ga-rich and As-rich conditions correspond to $\mu_{\rm Ga} = 0$ eV and $\mu_{\rm As} = 0$ eV, respectively. We also examine defect landscape in GaAs at a midpoint between the two extreme limits where $\mu_{\rm Ga} = \mu_{\rm As} = \Delta H ({\rm GaAs})/2$. The oxygen chemical potential ($\mu_{\rm O}$) is assumed to be limited by the formation of $\beta$-Ga$_2$O$_3$ ($\Delta H = -10.15$ eV at 0 K, compared to the experimental value of $-11.28$ eV under ambient conditions~\cite{Wagman1982}). Note that we are only interested in the trend in the formation energy among native defects or oxygen-related defects and in the defect transition level (which is independent of the choice of the atomic chemical potentials; see below) and thus make no attempt to find sets of $\mu_{i}$ values that can accurately reflect actual conditions under specific experimental situations.

From formation energies, one can calculate the {\it thermodynamic} transition level between charge states $q_1$ and $q_2$ of a defect, $\epsilon(q_1/q_2)$, defined as the Fermi-level position at which the formation energy of the defect in charge state $q_1$ is equal to that in charge state $q_2$~\cite{Freysoldt2014RMP}, i.e.,
\begin{equation}\label{eq;tl}
\epsilon(q_1/q_2) = \frac{E^f(X^{q_1}; \mu_e=0)-E^f(X^{q_2}; \mu_e=0)}{q_2 - q_1},
\end{equation}
where $E^f(X^{q}; \mu_e=0)$ is the formation energy of the defect X in charge state $q$ when the Fermi level is at the VBM ($\mu_e=0$). This $\epsilon(q_1/q_2)$ level [also referred to as the $(q_1/q_2)$ level], corresponding to a defect (energy) level, would be observed in, e.g., deep-level transient spectroscopy (DLTS) experiments where the defect in the final charge state $q_2$ fully relaxes to its equilibrium configuration after the transition. The {\it optical} transition level $E_{\rm opt}^{q_1/q_2}$ is, on the other hand, defined similarly but with the total energy of the final state $q_2$ calculated using the lattice configuration of the initial state $q_1$~\cite{Freysoldt2014RMP}. 

Note that spin-orbit coupling (SOC) has negligible effects on the thermodynamic and optical transition levels. Our computational tests performed on the substitutional oxygen impurity (O$_{\rm As}$) show that the inclusion of SOC changes the formation energy by less than 10 meV. Such effects cancel out in the transition levels, however.   

We calculate configuration coordinate diagrams for optical transitions and carrier capture coefficients and cross sections using the \textsc{nonrad} code \cite{Turiansky2021CPC} which implements the first-principles approach of Alkauskas et al.~\cite{Alkauskas2014PRB}. Effective electron masses are calculated from \textsc{vasp} data by using the post-processing \textsc{vaspkit} package \cite{Wang2021CPC}. The thermal emission rate of electrons, $e_n$, from select defect levels into the conduction band is also calculated using the following expression which relates $e_n$ to the electron capture cross section ($\sigma_n$)~\cite{Lang1974JAP,Wickramaratne2018APL}:
\begin{equation}\label{eq;en}
e_n(T) = \frac{\sigma_n(T)<v_n(T)>N_{\rm c}(T)}{g_{\rm c}}exp(-\frac{\Delta E}{k_{\rm B}T}),
\end{equation}
where $<v_n>$ is the average thermal velocity of electrons in the conduction band, $N_{\rm c}$ is the effective density of states of the conduction band, $g_{\rm c}$ is the valley degeneracy, and $\Delta E$ is the ionization energy of the defect.

\section{Results and discussion}\label{sec;results}

Let us begin with the basic bulk properties of the host material. In zincblende GaAs, each Ga is tetrahedrally coordinated with As; the Ga--As bond length is calculated to be 2.45 {\AA} and the lattice constant is 5.65 {\AA}. The material has the total static dielectric constant of 12.19. All these values are in agreement with experimental data~\cite{Blakemore1982JAP}. The VBM is predominantly composed of the As $4p$ states and a smaller contribution from the Ga $4p$ states and the CBM is predominantly the $4s$ states and a slightly smaller contribution from the As $4s$ states. The band gap is 1.51 eV (direct at $\Gamma$) at 0 K, which matches the experimental value at low temperatures \cite{Blakemore1982JAP}. The electron effective mass is calculated to be 0.089$m_0$, in reasonable agreement with experiments (0.067$m_0$)~\cite{Blakemore1982JAP}.

\subsection{Native defects}\label{sec;natives}

\begin{figure}
\vspace{0.2cm}
\includegraphics*[width=\linewidth]{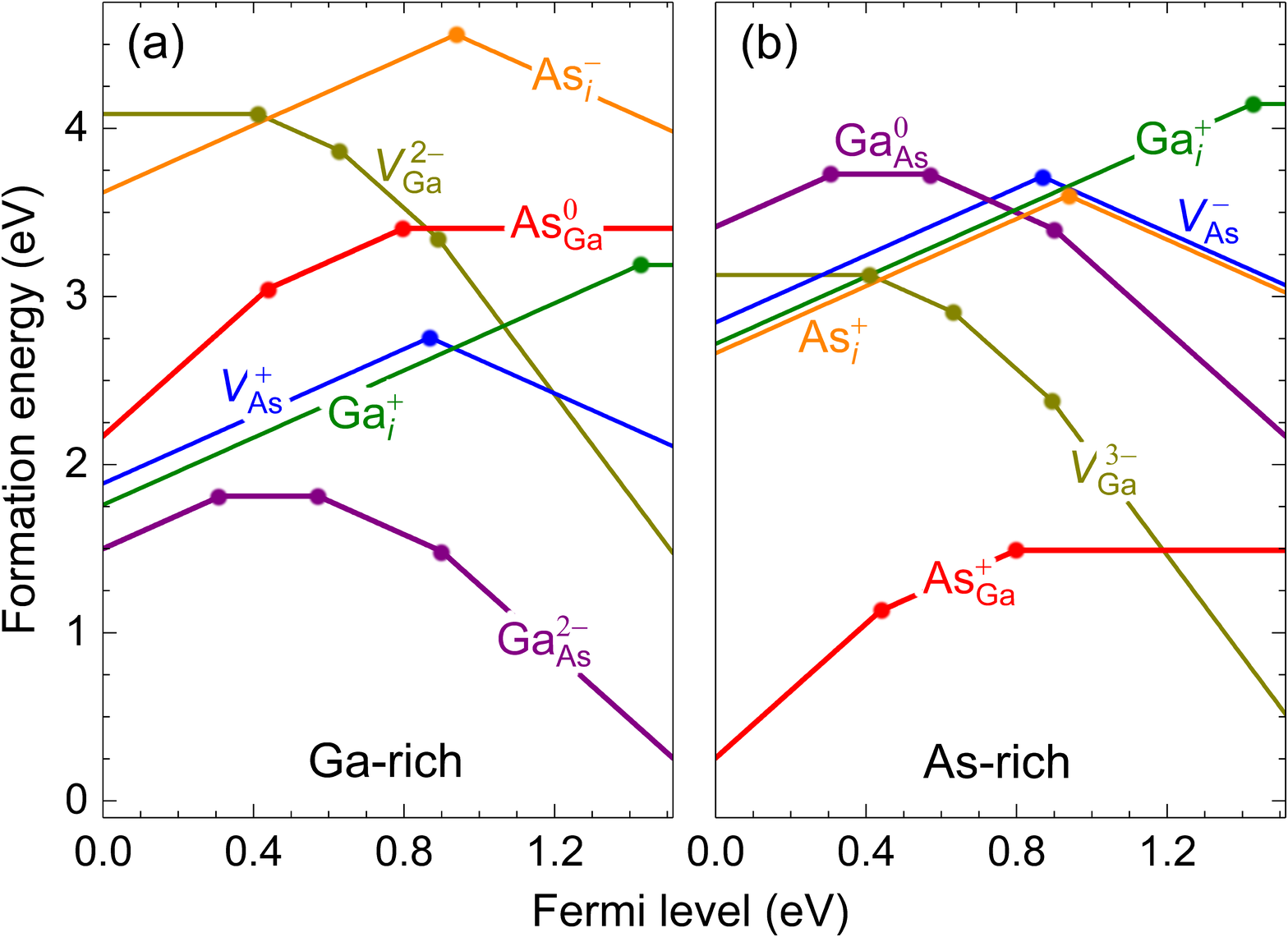}
\caption{Formation energies of native defects in GaAs, plotted as a function of the Fermi level from the VBM (at 0 eV) to the CBM (at 1.51 eV), under the extreme Ga-rich and As-rich conditions. For each defect, only segments corresponding to the lowest-energy charge states are shown. The slope indicates the charge state ($q$): positively (negatively) charged defects have positive (negative) slopes. Large solid dots connecting two segments mark the {\it defect level} $\epsilon(q_1/q_2)$.}
\label{fig;natives} 
\end{figure}

\begin{figure}
\vspace{0.2cm}
\includegraphics*[width=\linewidth]{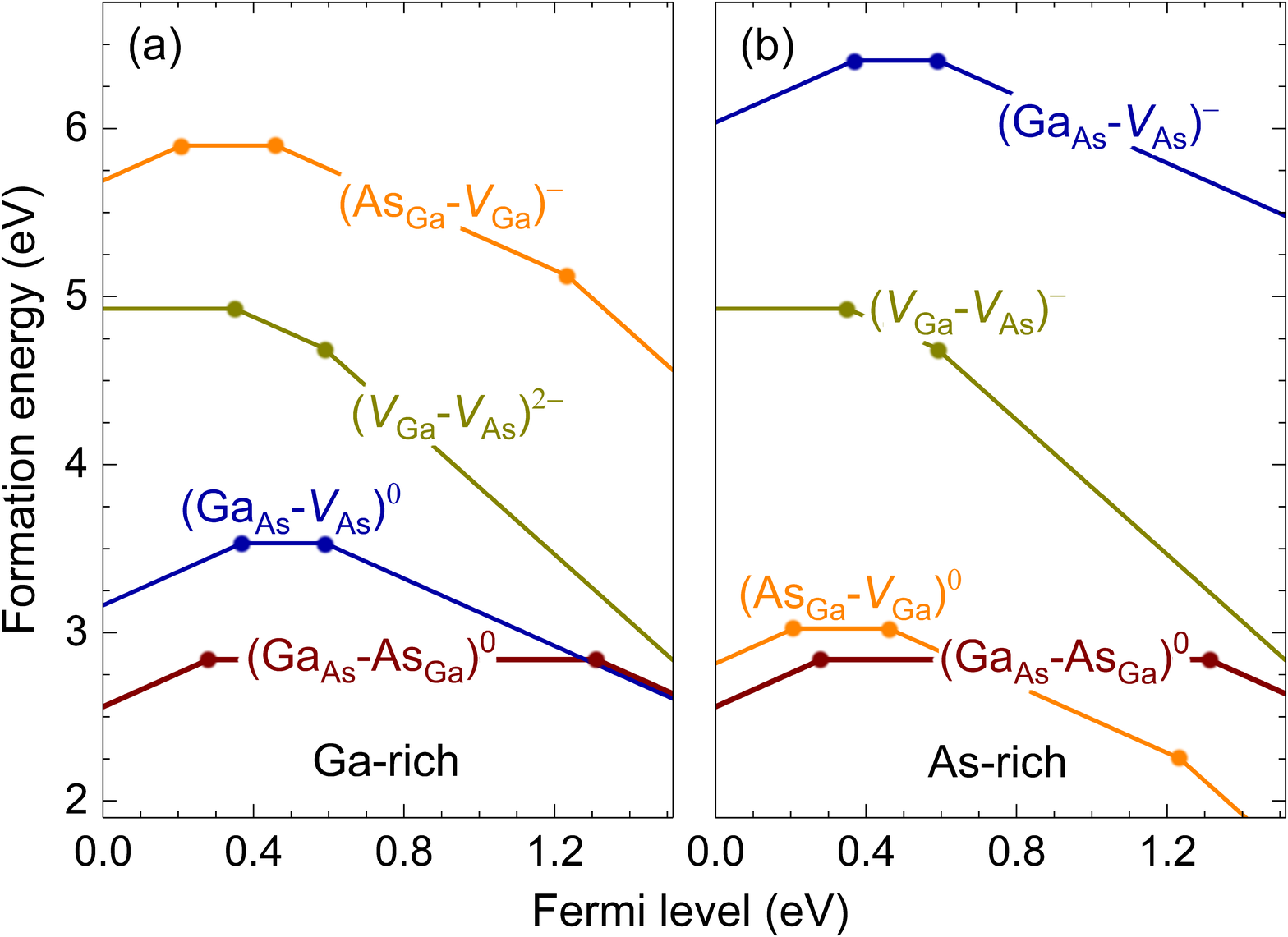}
\caption{Formation energies of native defect complexes in GaAs under the extreme Ga-rich and As-rich conditions.}
\label{fig;complexes} 
\end{figure}

Native defects in GaAs include Ga and As vacancies ($V_{\rm Ga}$ and $V_{\rm As}$), interstitials (Ga$_i$ and As$_i$), antisites (Ga$_{\rm As}$ and As$_{\rm Ga}$), and their complexes. Figures~\ref{fig;natives} and \ref{fig;complexes} show the formation energies of structurally, electronically, and energetically stable defect configurations in GaAs.   

In $V_{\rm Ga}$, the removal of a Ga atom to create the vacancy results in significant inward relaxation of the nearest As atoms, by 0.12 {\AA} (in the case of $V_{\rm Ga}^0$) to 0.27 {\AA} ($V_{\rm Ga}^{3-}$; see Fig.~\ref{fig;struct} in Appendix~\ref{sec;app}). The vacancy has another configuration, denoted as $V_{\rm Ga}^\ast$~\cite{Schultz2009MSMSE}, in which an adjacent As atom moves into the void created by $V_{\rm Ga}$ and leaves behind an As vacancy; see Fig.~\ref{fig;struct}. This configuration can thus be regarded as a complex of As$_{\rm Ga}$ (see below) and $V_{\rm As}$. The energy barrier is 1.05 eV for the $V_{\rm Ga}^0$ $\rightarrow$ $V_{\rm Ga}^{\ast,0}$ transition or 0.70 eV for $V_{\rm Ga}^{\ast,0}$ $\rightarrow$ $V_{\rm Ga}^0$, see Fig.~\ref{fig;em}, calculated by using the climbing-image nudged elastic-band (NEB) method \cite{ci-neb}; in the saddle-point configuration, the migrating As atom is about midway between the As and Ga sites. $V_{\rm Ga}$ is energetically more stable than $V_{\rm Ga}^\ast$ in the upper two-thirds of the band gap whereas $V_{\rm Ga}^\ast$ is more stable in the lower one-third, and the two configurations can pin the Fermi level at $\mu_e = 0.51$ eV if they are the dominant defects; see Fig.~\ref{fig;natives}. This level is in excellent agreement with the experimental value of 0.5--0.6 eV for the Fermi-level pinning position related to Ga vacancies formed at metal--GaAs contacts~\cite{Tersoff1986PRL,Walukiewicz1987JVSTB}.

In $V_{\rm As}$, an adjacent Ga atom is displaced toward the void left by the removal of the As atom, by from 0.46 {\AA} in $V_{\rm As}^+$ to 2.53 {\AA} in $V_{\rm As}^{3-}$ (In the latter, the Ga atom occupies the void at the As site). $V_{\rm As}$ is thus significantly off-center, and $V_{\rm As}^{3-}$ can be regarded as a complex of Ga$_{\rm As}^0$ (see below) and $V_{\rm Ga}^{3-}$; see Fig.~\ref{fig;struct}. The on-center $V_{\rm As}$ is higher in energy, by up to 1.17 eV in the $3-$ charge state. The transition from the on- to off-center $V_{\rm As}^{3-}$ is almost barrierless (the energy barrier is only 0.03 eV); Fig.~\ref{fig;em}.        

In Ga$_i$, the interstitial Ga atom is tetrahedrally coordinated with Ga, and octahedrally coordinated with As at a slightly longer distance. The configuration in which the interstitial Ga is tetrahedrally coordinated with nearest As neighbors, denoted as Ga$_i^{\ast}$, is higher in energy by 0.19 eV (0.20 eV) in the $0$ ($+$) charge state or lower in energy by 0.24 eV in the $2+$ charge state. The energy barrier is 1.21 eV for the Ga$_i^{0}$ $\rightarrow$ Ga$_i^{\ast,0}$ transition or 1.02 eV for Ga$_i^{\ast,0}$ $\rightarrow$ Ga$_i^{0}$; Fig.~\ref{fig;em}. In As$_i^+$, the interstitial As atom is close to one Ga atom and two As atoms, whereas As$_i^-$ forms an As--As split structure, see Fig.~\ref{fig;struct}.

In Ga$_{\rm As}$, the antisite Ga atom is tetrahedrally coordinated with Ga and slightly off-center with the displacement varies from 0 {\AA} in Ga$_{\rm As}^{2-}$ (where the antisite Ga atom is four-fold coordinated; see Fig.~\ref{fig;struct}) to 0.28 {\AA} in Ga$_{\rm As}^+$ (three-fold coordinated). In As$_{\rm Ga}$, the antisite As atom is tetrahedrally coordinated with As; see Fig.~\ref{fig;struct}. As$_{\rm Ga}^+$ (spin $S=1/2$) is paramagnetic with one unpaired electron. In another configuration, denoted as As$_{\rm Ga}^\ast$~\cite{Schultz2009MSMSE}, the antisite As atom is displaced away from one of the nearest As neighbors and becomes three-fold coordinated with As; it is off-center by 1.37 {\AA} in As$_{\rm Ga}^{\ast,0}$, see Fig.~\ref{fig;struct}. As$_{\rm Ga}^{\ast,0}$ is higher in energy than As$_{\rm Ga}^0$ by 0.38 eV. The energy barrier is 0.76 eV for the As$_{\rm Ga}^0$ $\rightarrow$ As$_{\rm Ga}^{\ast,0}$ transition or 0.38 eV for As$_{\rm Ga}^{\ast,0}$ $\rightarrow$ As$_{\rm Ga}^0$, see Fig.~\ref{fig;em}, in agreement with the experimental value of 0.34 eV~\cite{Mitonneau1979SSC,Vincent1982JAP}.

These isolated native defects have been studied and reported extensively by many authors in the past decades. Readers are referred to, e.g., Schultz and Anatole von Lilienfeld~\cite{Schultz2009MSMSE} for a more comprehensive discussion.     

The antisite pair, Ga$_{\rm As}$-As$_{\rm Ga}$, is formed by switching the positions of one Ga and one As that are nearest neighbors to each other. In (Ga$_{\rm As}$-As$_{\rm Ga}$)$^0$, for example, the distance between the two defects is 2.49 {\AA}. The Schottky defect pair (divacancy), $V_{\rm Ga}$-$V_{\rm As}$, is formed by removing a Ga-As pair from the lattice. Complexes between an antisite and a vacancy include Ga$_{\rm As}$-$V_{\rm As}$ and As$_{\rm Ga}$-$V_{\rm Ga}$. Unlike some of the native defects discussed above, the constituents in these defect complexes are only slightly off-center; see Fig.~\ref{fig;struct}. The binding energy of these complexes in different charge states with respect to their constituents and the ground-state spin of all isolated defect and defect complex configurations are reported in Table~\ref{tab;complex}. One should note that, as discussed in Ref.~\citenum{walle:3851}, in order for a defect complex to have a higher concentration than its constituents under thermodynamic equilibrium, the binding energy needs to be greater than the larger of the formation energies of the isolated constituents. 

\begin{table}
\caption{Defect energy levels (in eV, with respect to the VBM, $E_{\mathrm{v}}$) induced by native defects and impurities.}\label{tab;defectlevel}
\begin{center}
\begin{ruledtabular}
\begin{tabular}{ll}
Defect & Defect levels$^{a}$ \\
\colrule
$V_{\rm Ga}$ & $(0/-) = 0.41$, $(-/2-) = 0.63$, \\
& $(2-/3-) = 0.89$ \\
$V_{\rm Ga}^\ast$ & $(3+/+) = 0.36$, $(+/0) = 0.93$, $(0/-) = 1.46$;\\
& $(3+/2+) = 0.52$, $(2+/+) = 0.20$ \\
$V_{\rm As}$ & $(+/-) = 0.89$, $(-/3-) = 1.28$;\\
& $(+/0) = 1.06$, $(0/-) = 0.73$, $(2-/3-) = 0.96$ \\
Ga$_{\rm As}$ & $(+/0) = 0.31$, $(0/-) = 0.58$, $(-/2-) = 0.89$ \\
As$_{\rm Ga}$ & $(2+/+) = 0.44$, $(+/0) = 0.80$ \\
As$_{\rm Ga}^\ast$ & $(+/0) = 0.02$ \\
Ga$_{i}$  & $(+/0) = 1.43$ \\
Ga$_{i}^\ast$  & $(2+/+) = 0.27$, $(+/0) = 1.42$ \\
As$_i$ & $(+/-) = 0.94$; $(+/0) = 0.91$, $(0/-) = 0.96$  \\
Ga$_{\rm As}$-As$_{\rm Ga}$ & $(+/0) = 0.28$, $(0/-) = 1.31$ \\
As$_{\rm Ga}$-$V_{\rm Ga}$ & $(+/0) = 0.21$, $(0/-) = 0.46$, $(-/2-) = 1.23$ \\
Ga$_{\rm As}$-$V_{\rm As}$ & $(+/0) = 0.37$, $(0/-) = 0.59$  \\
$V_{\rm Ga}$-$V_{\rm As}$ & $(0/-) = 0.35$, $(-/2-) = 0.59$ \\
O$_{\rm As}$ & $(+/-) = 0.99$; $(+/0) = 0.99$, $(0/-) = 0.99$ \\
O$_{i,{\rm tet}}$ & $(+/0) = 0.05$, $(0/-) = 0.19$, $(-/2-) = 0.30$ \\
O$_{\rm As}$-As$_{\rm Ga}$ & $(+/0) = 1.27$, $(0/-) = 1.35$ \\
O$_{\rm As}$-2As$_{\rm Ga}$ & $(+/-) = 1.15$; $(+/0) = 1.27$, $(0/-) = 1.03$ \\
\end{tabular}
\end{ruledtabular}
\end{center}
\begin{flushleft}
 $^a$Electronically stable but energetically metastable defect levels, if available, are listed after the semicolon.
\end{flushleft}
\end{table}

All these defects are electrically active and induce energy levels in the band gap of the GaAs host as summarized in Table~\ref{tab;defectlevel}. Energetically, As$_{\rm Ga}$ has the lowest formation energy under the extreme As-rich condition, except near the CBM where $V_{\rm Ga}^{3-}$ is lowest in energy; Ga$_{\rm As}$ has the lowest formation energy in the entire range of Fermi-level values under the extreme Ga-rich condition; see Fig.~\ref{fig;natives}. More realistic experimental conditions are between the extreme Ga-rich and As-rich limits. At the midpoint between the two limits where $\mu_{\rm Ga}$ = $\mu_{\rm As}$, for example, Ga$_{\rm As}$ and As$_{\rm Ga}$ are the dominant native defects and pin the Fermi level near midgap where Ga$_{\rm As}^-$ and As$_{\rm Ga}^+$ have equal formation energies; see Fig.~\ref{fig;mid}(a). This is also consistent with the fact that Ga$_{\rm As}$-As$_{\rm Ga}$ has the lowest formation energy in the entire range of Fermi-level values under the same condition; Fig.~\ref{fig;mid}(b). 

The As-rich (or As-excess) condition often used in experiments should correspond to a point near the extreme As-rich limit. In this case, and especially under $n$-type conditions and in the absence of negatively charged (intentional or unintentional) impurities that have a lower formation energy than $V_{\rm Ga}$, As$_{\rm Ga}$ is charge compensated by $V_{\rm Ga}$. This is thus consistent with experiments showing that As$_{\rm Ga}$ is abundant in As-excess GaAs samples and the antisite defect is charge compensated by $V_{\rm Ga}$~\cite{Bliss1992JAP,Liu1994APL,Luysberg1998JAP}. Other isolated native defects have high formation energies, see Fig.~\ref{fig;natives}, and may thus occur with sizeable concentrations only under non-equilibrium conditions.  

On the basis of the calculated defect levels, As$_{\rm Ga}$ can be identified with the $EL2$ center with the energies at 0.50--0.54 eV and 0.75--0.77 eV above the VBM observed in experiments~\cite{Martin1977EL,Weber1982JAP,Lagowski1985APL,Omling1988PRB,Spicer1988JVST} with As$_{\rm Ga}^\ast$ as its metastable configuration as proposed in Refs.~\citenum{Dabrowski1988PRL} and \citenum{Chadi1988PRL}. The results for the $(2+/+)$ and $(+/0)$ levels of As$_{\rm Ga}$, at 0.44 and 0.80 eV (Table~\ref{tab;defectlevel}), respectively, obtained in our HSE calculations are in agreement with those calculated at the $G_0W_0$@HSE level reported by Chen and Pasquarello~\cite{Chen2017PRB} in which the atomic structure was relaxed at the HSE level (This work also highlights the need to perform structural relaxations at the HSE level). Similar numbers have also been reported by other authors~\cite{Schultz2009MSMSE,Komsa2012JPCM,Wright2015PRB,Fluckey2026CMS}.

The $(-/3-)$ level of $V_{\rm As}$ at 0.23 eV below the CBM may seem to be related to the $E3$ (or $S3$) center level observed at 0.30 eV~\cite{Pons1985JPC,Bourgoin1988JAP,Taghizadeh2022MSSP}. However, in such a negative-$U$ center, DLTS does not directly measure the $(-/3-)$ level~\cite{Wickramaratne2018APL} but likely the $(2-/3-)$ level at $E_{\mathrm{c}}-0.56$ eV as the other single charge transition level, $(-/2-)$, is above the CBM. This thus does not seem to support the assignment of $V_{\rm As}$ to $E3$. We also find that $V_{\rm Ga}$-$V_{\rm As}$ does not induce any defect levels right below the CBM, see Fig.~\ref{fig;natives}, which is different from the LDA/GGA results of Schultz and Anatole von Lilienfeld~\cite{Schultz2009MSMSE} who assigned the divacancy to the $E1$ and $E2$ centers. Our results are, therefore, inconclusive about the origin of the $E1$--$E3$ centers. Finally, the fact that Ga$_i$ is stable in the $0$ and $+$ states is consistent with the experimental observation in which zinc diffusion in GaAs was found to be mediated by neutral and singly positively charged Ga$_i$~\cite{Bracht2005PRB}.     

\begin{figure}
\vspace{0.2cm}
\includegraphics*[width=\linewidth]{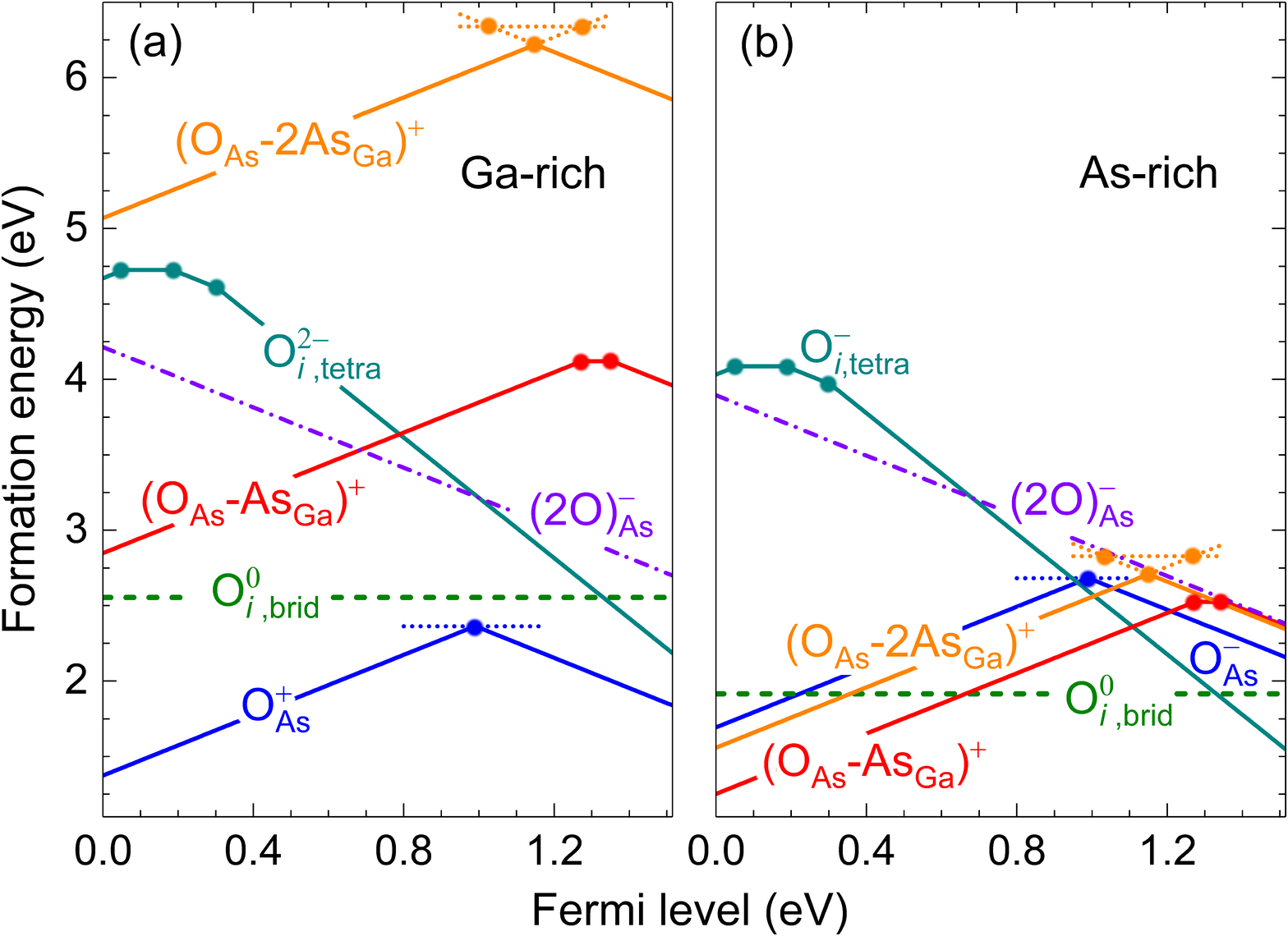}
\caption{Formation energies of O-related defects in GaAs under the extreme Ga-rich and As-rich conditions. The formation energies of O$_{\rm As}^0$ and (O$_{\rm As}$-2As$_{\rm Ga}$)$^0$ are represented by dotted blue and orange horizontal segments, respectively.}
\label{fig;oxygen} 
\end{figure}

\subsection{Oxygen impurities}\label{sec;oxygen}

Figure \ref{fig;oxygen} shows the formation energies of O-related defects in GaAs. O$_{\rm As}$ is electronically stable as O$_{\rm As}^+$, O$_{\rm As}^0$, and O$_{\rm As}^-$ but O$_{\rm As}^0$ is energetically metastable; the $(+/-)$, $(+/0)$, and $(0/-)$ levels are degenerate in energy; see also Table~\ref{tab;defectlevel}. The substitutional O atom is significantly off-center and forms approximately an OGa$_3$ trigonal planar (but stays slightly above the Ga triangle plane) with the O--Ga distance of about 2 {\AA}; see Fig.~\ref{fig;struct} in Appendix A. O$_{\rm As}$ can thus be referred to as being quasi-substitutional. 

Oxygen can also be incorporated into GaAs at interstitial sites. In the O$_{i,{\rm brid}}$ configuration, the interstitial atom bridges one Ga and one As and forms a puckered bond-center Ga--O--As defect. It pushes the two host atoms slightly off-center; the O--Ga and O--As distances are 1.83 {\AA} and 1.75 {\AA}, respectively. In the O$_{i,{\rm tet}}$ configuration, the oxygen is tetrahedrally coordinated with Ga; the O--Ga distance is 2.05 {\AA} in O$_{i,{\rm tet}}^{2-}$; see Fig.~\ref{fig;struct}. 

Other O-related defects include (2O)$_{\rm As}$ in which an As atom is substituted by two O atoms. (2O)$_{\rm As}$ forms a double-wing shape with two OGa$_3$ (approximately) trigonal planars sharing two Ga atoms. Finally, O$_{\rm As}$-As$_{\rm Ga}$ and O$_{\rm As}$-2As$_{\rm Ga}$ are defect complexes consisting of O$_{\rm As}$ and one and two As$_{\rm Ga}$ defects, respectively. In (O$_{\rm As}$-As$_{\rm Ga}$)$^+$ and (O$_{\rm As}$-As$_{\rm Ga}$)$^0$, the local structure of O$_{\rm As}$ remains largely intact with O forming three O--Ga bonds with bond lengths of 1.98--1.99 {\AA}, see Fig.~\ref{fig;struct}; whereas in (O$_{\rm As}$-As$_{\rm Ga}$)$^-$ the oxygen atom forms two O--Ga bonds, i.e., a puckered bond-center Ga--O--Ga structure, with the O--Ga bond length of 1.87 {\AA}. In all three charge states of O$_{\rm As}$-2As$_{\rm Ga}$, O forms a puckered bond-center Ga--O--Ga structure with O--Ga bond lengths of 1.83--1.84 {\AA}; see Fig.~\ref{fig;struct}. The electronically stable charge states of these complexes and the binding energy with respect to their isolated constituents are reported in Table S1.

Except for O$_{i,{\rm brid}}$ (stable only as electrically inactive O$_{i,{\rm brid}}^0$) and (2O)$_{\rm As}$ [stable only as (2O)$_{\rm As}^-$, a shallow acceptor], other O-related defects all induce energy levels in the band gap of the host; see Table~\ref{tab;defectlevel}. Among the O-related defects, O$_{\rm As}$ and O$_{\rm As}$-2As$_{\rm Ga}$ both have a thermodynamically metastable middle (neutral) charge state which is paramagnetic with one unpaired electron ($S=1/2$). The latter is a negative-$U$ center with $U$ $=$ $\epsilon(+/0) - \epsilon(0/-)$ $=$ $0.24$ eV; the former has $U = 0$ eV. 

Energetically, O$_{\rm As}$ is found to be the lowest-energy O-related defect in the entire range of Fermi-level values under the extreme Ga-rich condition; see Fig.~\ref{fig;oxygen}(a). Under the extreme As-rich condition, (O$_{\rm As}$-As$_{\rm Ga}$)$^+$, O$_{i,{\rm brid}}^0$, or O$_{i,{\rm tet}}^{2-}$ is the lowest-energy O-related defect, depending the position of the Fermi level; Fig.~\ref{fig;oxygen}(b). At the midpoint between the two limits, O$_{\rm As}^+$, O$_{i,{\rm brid}}^0$, O$_{\rm As}^-$, or O$_{i,{\rm tet}}^{2-}$ is the lowest-energy defect; see Fig.~\ref{fig;mid}(c). Overall, under As-rich conditions, the O-related defects are relatively close in formation energies. Multiple defect centers, including isolated defects and defect complexes, are thus expected to exist in GaAs prepared under such conditions.

\begin{figure}
\vspace{0.2cm}
\includegraphics*[width=\linewidth]{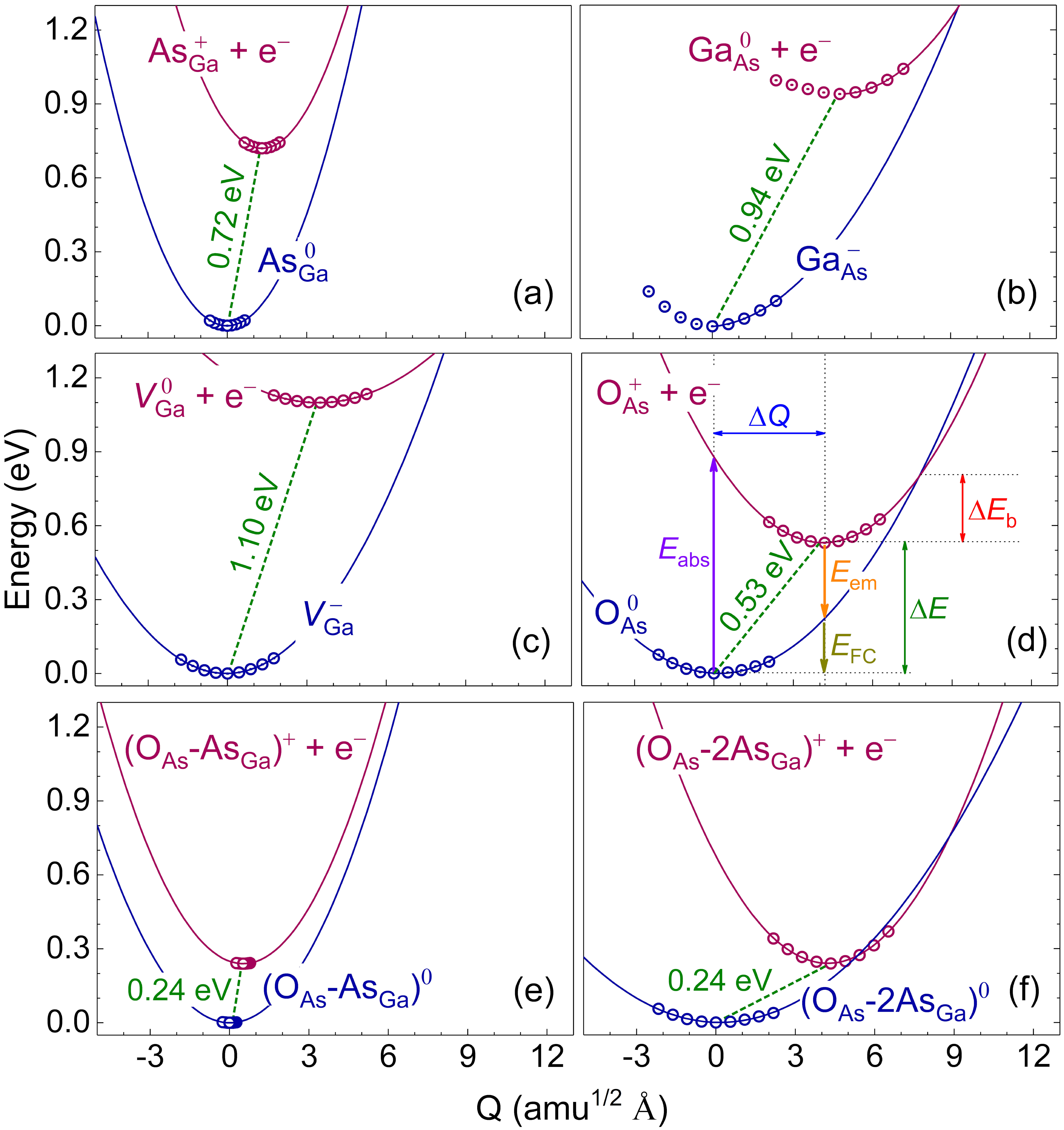}
\caption{Configuration coordinate diagrams for relevant optical transitions in GaAs. As illustrated in the case of the ${\rm O}_{\rm As}^+ + e^- \rightleftharpoons {\rm O}_{\rm As}^0$ transitions, $\Delta E$ is the ionization energy (i.e., ZPL), $E_{\rm abs}$ ($E_{\rm em}$) is the peak absorption (emission) energy, $E_{\rm FC}$ is the Franck-Condon shift, $\Delta Q$ is the mass-weighted difference between the geometries of the excited and ground states, and $\Delta E_{\rm b}$ is the electron capture barrier.}
\label{fig;cc} 
\end{figure}

Experimentally, multiple O-related defect centers have indeed been observed in GaAs~\cite{Skowronski1991JAP,Skowronski1992PRB}. Those include the electrically inactive O$_{i,{\rm brid}}$~\cite{Schneider1989APL,Skowronski1991APL,Skowronski1991JAP,Skowronski1992PRB}. Another center (sometimes referred to as ``OX'' or ``O$_{\rm oc}$'') has been characterized as being a quasi-substitutional oxygen where the oxygen forms the Ga--O--Ga configuration and which exhibits the negative-$U$ character with a paramagnetic metastable charge state~\cite{Zhong1988APL,Schneider1989APL,Alt1989APLoxygen,Alt1990PRL,Song1990JAP,Skowronski1990APL,Skowronski1991APL,Neild1991APL,Skowronski1991JAP,Skowronski1992PRB,Jordan1992SST,Linde1995APL,Alt2007JAP}. Such a description points to O$_{\rm As}$-2As$_{\rm Ga}$ with the Ga--O--Ga structure we described earlier. Note that in a DLTS measurement, in principle, there are two emission processes that correspond to the thermal emission from the $(+/0)$ and $(0/-)$ levels of O$_{\rm As}$-2As$_{\rm Ga}$. In practice, only one emission process--which is the slowest of the two processes--that would be seen in experiment leading to a single peak in the capacitance transient~\cite{Wickramaratne2018APL}. As will be shown in the next section, that process is associated with the $(0/-)$ level (at 1.03 eV above the VBM or $E_{\rm c} - 0.48$ eV; see Table~\ref{tab;defectlevel}) which has a lower emission rate than the $(+/0)$ level (at $E_{\rm c} - 0.24$ eV). Our results for the $(0/-)$ level is in reasonable agreement with the activation energy of 0.55 eV obtained from DLTS measurements reported for ``OX''~\cite{Neild1991APL}. The calculated $(+/-)$ level (at $E_{\rm c} - 0.36$ eV) is, on the other hand, comparable to the Fermi-level pinning position at 0.43 eV~\cite{Alt1989APLoxygen} or 0.36 eV~\cite{Alt2007JAP} observed in Hall measurements. Our results thus support the O$_{\rm As}$-2As$_{\rm Ga}$ model for ``OX'' proposed by Pesola et al.~\cite{Pesola1999PRB}. Note, however, that their $(+/-)$ level is at $E_{\rm c}-0.83$ eV.

In general, our results for the O-related defects are often qualitatively or quantitatively different from those reported in the literature. Colleoni and Pasquarello~\cite{Colleoni2016PRB}, for example, found O$_{\rm As}^+$ and O$_{\rm As}^0$ to be at the As site whereas O in O$_{\rm As}^-$ is off-center and forms two O--Ga bonds, and O$_{\rm As}$ was found to be a positive-$U$ center, which is in contrast to our findings. It was also concluded that O$_{i,{\rm tet}}$ (referred to as ``O$_i$-Ga$_4$'' in Ref.~\citenum{Colleoni2016PRB}) is stable in the neural charge state only and has a much higher energy than O$_{i,{\rm brid}}$, whereas we find that the defect is stable in multiple charge states and can be the lowest-energy O-related defect under $n$-type conditions; Fig.~\ref{fig;oxygen}(b). The discrepancies can be ascribed to the smaller supercell size, the PBE-relaxed structures, and other parameters and procedures used in the previous HSE calculations. 

\begin{figure}
\vspace{0.2cm}
\includegraphics*[width=\linewidth]{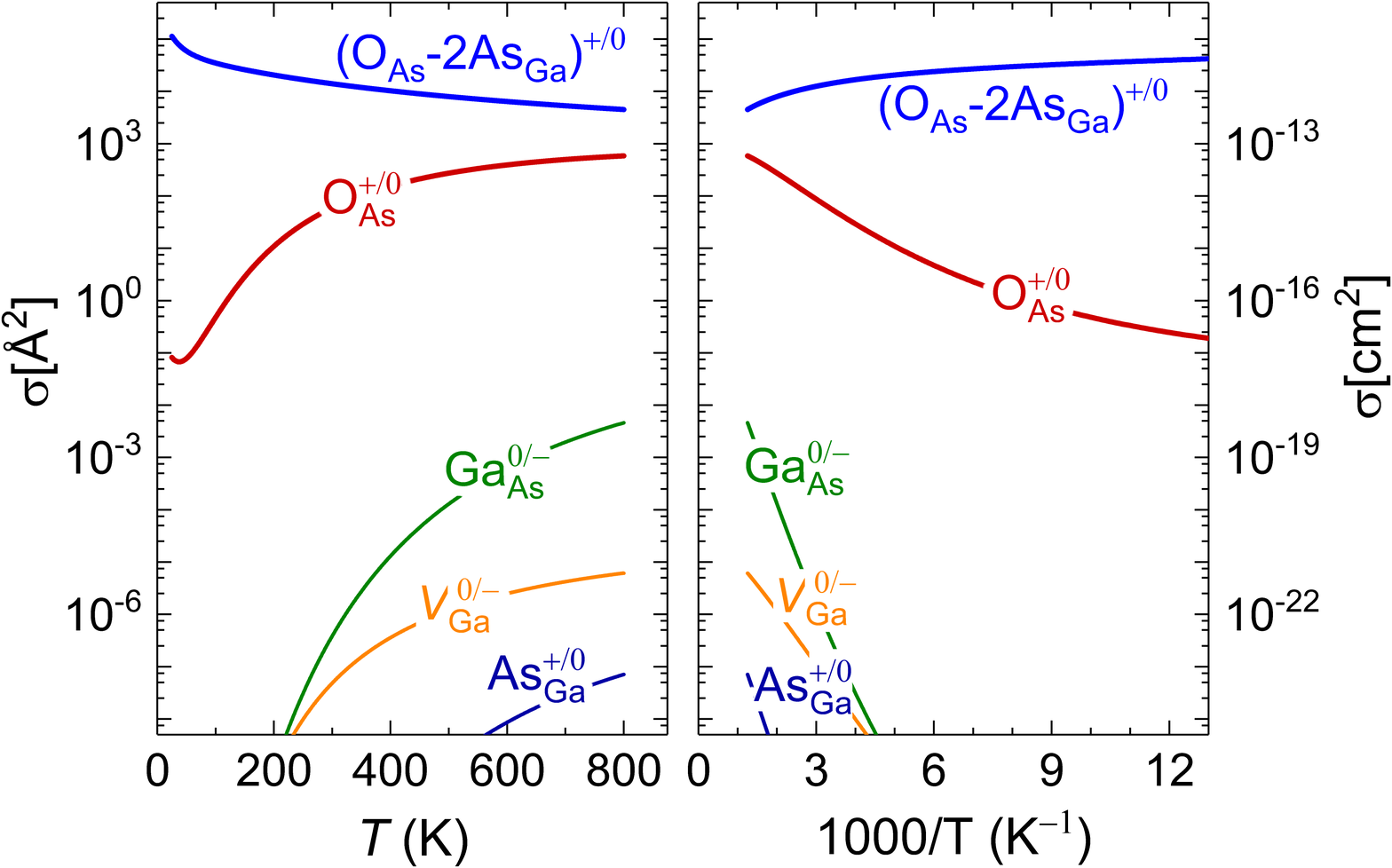}
\caption{Nonradiative electron capture cross sections (in {\AA}$^2$ and cm$^2$) of relevant defect centers in GaAs.}
\label{fig;sigma} 
\end{figure}

\subsection{Optical properties}

\begin{table*}
\caption{Key parameters associated with the optical transitions considered in this work; see text.}\label{tab;opt}
\begin{center}
\begin{ruledtabular}
\begin{tabular}{lccccccccc}
Optical transition & $\Delta E$ & $E_{\rm abs}$ & $E_{\rm em}$ & $E_{\rm FC}$ & $\Delta Q$ & $\hbar \Omega_{\rm e}$ & $\hbar \Omega_{\rm g}$ & $S_{\rm g}$ &  $\tilde{W}_{\rm eg}$\\
\colrule
${\rm As}_{\rm Ga}^+ + e^- \rightleftharpoons {\rm As}_{\rm Ga}^0$ & 0.72 &0.81 &0.63 &0.09 & 1.31 & 21.43 & 20.51 &4.39 & 3.73$\times$10$^{-2}$ \\
${\rm Ga}_{\rm As}^0 + e^- \rightleftharpoons {\rm Ga}_{\rm As}^-$ & 0.94 &1.13 &0.72 &0.22 & 2.78 & 14.03 & 16.19 &13.59 & 2.82$\times$10$^{-4}$ \\
$V_{\rm Ga}^0 + e^- \rightleftharpoons V_{\rm Ga}^-$ & 1.10 &1.23 &0.85 &0.25 & 3.50 & 9.35 & 12.68 &19.72 & 2.09$\times$10$^{-5}$ \\
${\rm O}_{\rm As}^+ + e^- \rightleftharpoons {\rm O}_{\rm As}^0$ & 0.53 &0.85 &0.40 &0.13 & 4.18 & 13.06 & 10.80 &12.04 & 4.26$\times$10$^{-2}$ \\
(${\rm O}_{\rm As}$-${\rm As}_{\rm Ga}$)$^+$ $+$ $e^-$ $\rightleftharpoons$ (${\rm O}_{\rm As}$-${\rm As}_{\rm Ga}$)$^0$ & 0.24 &0.25 &0.23 &0.01 & 0.52 & 17.49 & 16.34 &0.61 & 3.51$\times$10$^{-2}$ \\
(${\rm O}_{\rm As}$-$2{\rm As}_{\rm Ga}$)$^+$ $+$ $e^-$ $\rightleftharpoons$ (${\rm O}_{\rm As}$-$2{\rm As}_{\rm Ga}$)$^0$ & 0.24 &0.61 &0.16 &0.08 & 4.34 & 14.25 & 9.14 &8.75 & 4.66$\times$10$^{-2}$  \\
(${\rm O}_{\rm As}$-$2{\rm As}_{\rm Ga}$)$^0$ $+$ $e^-$ $\rightleftharpoons$ (${\rm O}_{\rm As}$-$2{\rm As}_{\rm Ga}$)$^-$ & 0.48 &0.83 &0.11 &0.37 & 4.77 & 10.50 & 16.49 &22.44 & 3.29$\times$10$^{-2}$  \\
\end{tabular}
\end{ruledtabular}
\end{center}
\end{table*}

Here, we focus on As$_{\rm Ga}$, Ga$_{\rm As}$, $V_{\rm Ga}$, O$_{\rm As}$, O$_{\rm As}$-As$_{\rm Ga}$, and O$_{\rm As}$-2As$_{\rm Ga}$ as they possess defect levels in the band gap and have relatively low energies (and thus can occur with a sizable concentration) when GaAs is prepared under As-rich conditions. Figure~\ref{fig;cc} shows the configuration coordinate diagram for relevant optical transitions, constructed using the \textsc{nonrad} code~\cite{Turiansky2021CPC}. It expresses the total energy as a function of the generalized coordinate $Q$ which captures collective atomic displacements in an one-dimensional approximation by linear interpolation between geometries of ground and excited states~\cite{Alkauskas2014PRB}. Such diagrams are useful for the examination of both defect-assisted radiative and nonradiative processes.

We begin with possible defect-to-band radiative transitions. Under p-type conditions, i.e., when the Fermi level is below midgap, O$_{\rm As}$ is stable as O$_{\rm As}^+$; see Fig.~\ref{fig;natives}(b). This defect configuration can capture an electron from the CBM (e.g., previously excited to from the valence to the conduction band) and emits a photon. The peak emission energy ($E_{\rm em}$) corresponds to the optical transition level $E_{\rm opt}^{+/0}$, i.e., the energy difference between O$_{\rm As}^+$ and O$_{\rm As}^0$ in the lattice geometry of O$_{\rm As}^+$. As shown in Fig.~\ref{fig;cc}(d) and Table~\ref{tab;opt}, this emission peak is at 0.40 eV, with a relaxation energy (i.e., the Franck-Condon shift) of $0.13$ eV [$E_{\rm FC} = E_{\rm em} - \Delta E$, where $\Delta E$ is the ionization energy or the zero-phonon line (ZPL) energy which, for electron capture, is the transition level $\epsilon(q_1/q_2)$ referenced to the CBM]. We also consider the absorption process. In the case of O$_{\rm As}$, an electron can be excited from O$_{\rm As}^0$ to the conduction band, with a peak absorption energy ($E_{\rm abs}$) of 0.85 eV, corresponding to $E_{\rm opt}^{0/+}$, given by the energy difference between O$_{\rm As}^0$ and O$_{\rm As}^+$, both in the geometry of O$_{\rm As}^0$. Similar calculations are carried out for As$_{\rm Ga}$, Ga$_{\rm As}$, $V_{\rm Ga}$, O$_{\rm As}$-As$_{\rm Ga}$, and O$_{\rm As}$-2As$_{\rm Ga}$ and results are reported in Table~\ref{tab;opt}. We find that the emission and absorption energies are all in the near infrared range. The (${\rm O}_{\rm As}$-$2{\rm As}_{\rm Ga}$)$^0$ $\rightarrow$ (${\rm O}_{\rm As}$-$2{\rm As}_{\rm Ga}$)$^+$ $+$ $e^-$ transition with the absorption energy of 0.61 eV, see Fig.~\ref{fig;cc}(f) and Table~\ref{tab;opt}, may correspond to the so-called $B \rightarrow B'$ transition with the energy of 0.60--0.65 eV in Alt's experiment~\cite{Alt1990PRL}.    

Next, we calculate the nonradiative electron capture cross section at the selected defects. Table~\ref{tab;opt} shows key parameters extracted from \textsc{nonrad} calculations, including mass-weighted differences between the geometries of the excited (e) and ground (g) states ($\Delta Q$, in amu$^{1/2}${\AA}), energies of the effective vibrations ($\hbar \Omega_{\{\rm e,g\}}$, meV), Huang-Rhys factors for the ground state ($S_{\rm g} = E_{\rm FC}/\hbar\Omega_{\rm g}$), and electron-phonon coupling matrix elements ($\tilde{W}_{\rm eg}$, eV/amu$^{1/2}${\AA})~\cite{Alkauskas2014PRB,Turiansky2021CPC}. The harmonic approximation is assumed in the fitting of the potential energy surfaces. The energy curve for Ga$_{\rm As}$ is rather anharmonic, however. Following the argumentation in Ref.~\citenum{Alkauskas2014PRB}, we perform the parabolic fit for Ga$_{\rm As}^-$ (Ga$_{\rm As}^0$) only for $Q>0$ ($Q>\Delta Q$), thus focusing only on the range of $Q$ values where the energy curves cross; see Fig.~\ref{fig;cc}(b).

Figure~\ref{fig;sigma} shows the nonradiative electron capture cross sections ($\sigma_n$). Among the defects under consideration, we find that O$_{\rm As}$ and O$_{\rm As}$-2As$_{\rm Ga}$ have high $\sigma_n$ values, about 6.0$\times$10$^{-15}$ cm$^2$ at the O$_{\rm As}^{+/0}$ level, 1.4$\times$10$^{-12}$ cm$^2$ at the (O$_{\rm As}$-2As$_{\rm Ga}$)$^{+/0}$ level, or 1.9$\times$10$^{-14}$ cm$^2$ at the (O$_{\rm As}$-2As$_{\rm Ga}$)$^{0/-}$ level, all obtained at room temperature. These values are within an order of magnitude of the range of experimental values for the dominant defect centers in GaAs reported in the literature~\cite{Henry1977PRB,Martin1977EL,Mitonneau1979RPA,Neild1991APL}. The nonradiative electron capture cross section at As$_{\rm Ga}$, Ga$_{\rm As}$, and $V_{\rm Ga}$ is much lower and negligible; $\sigma_n$ at O$_{\rm As}$-As$_{\rm Ga}$ is even lower than that at As$_{\rm Ga}$ and is thus not included in Fig.~\ref{fig;sigma}. The high capture cross section at O$_{\rm As}$ or O$_{\rm As}$-2As$_{\rm Ga}$ is resulted from the low electron capture barrier $\Delta E_{\rm b}$ (in a semiclassical sense), as seen in Fig.~\ref{fig;cc}, and the relatively high electron-phonon coupling matrix element, see Table~\ref{tab;opt}. They can act as effective carrier traps or recombination centers which reduce carrier lifetime and mobility and thus may hurt (or help) the performance of GaAs-based devices (depending on specific applications). As$_{\rm Ga}$ (like O$_{\rm As}$-As$_{\rm Ga}$), on the other hand, has a negligible electron capture cross section (3.5$\times$10$^{-28}$ cm$^2$) mainly because it has a very high electron capture barrier ($\Delta E_{\rm b} \sim \infty$); see Fig.~\ref{fig;cc}.  With such a small electron capture cross section (and thus coefficient), and even if $EL2$ has a concentration that is several orders of magnitude higher than O$_{\rm As}$ or O$_{\rm As}$-2As$_{\rm Ga}$, the capture rate at $EL2$ is still negligible compared to that at the O$_{\rm As}$-related centers. This indicates that the isolated As$_{\rm Ga}$ cannot be the ``main electron trap'' in GaAs as commonly believed~\cite{Bourgoin2001SST,Martin1977EL,Mitonneau1979RPA,Weber1982JAP}. The observed electron trapping would thus be caused by other centers such as O$_{\rm As}$ or O$_{\rm As}$-2As$_{\rm Ga}$ that may co-exist with the antisite.

Finally, to help identify the DLTS level associated with the negative-$U$ ``OX'' center discussed earlier, we calculate the electron emission, $e_n$, from the $(+/0)$ and $(0/-)$ levels of O$_{\rm As}$-2As$_{\rm Ga}$ into the conduction band. Figure~\ref{fig;emission} shows that the emission rate at the $(0/-)$ level is lower than that at the $(+/0)$ level by several orders of magnitude. This indicates that in a conventional DLTS measurement the slower process associated with the $(0/-)$ level would be seen in the capacitance transient.   

\section{Conclusions} 

We have used a combination of structural, electronic, and optical characterizations to better understand native defects and oxygen impurities in GaAs, especially the As antisite and O-related defects. Among the native defects, Ga$_{\rm As}$, As$_{\rm Ga}$, and/or $V_{\rm Ga}$ are found to be the dominant native defects under thermodynamic equilibrium in which As$_{\rm Ga}$ and $V_{\rm Ga}$ are charge-compensating defects under As-rich conditions. The isolated antisite As$_{\rm Ga}$ can be identified with the $EL2$ center; however, it has a negligible electron capture cross section and thus cannot be an electron trap. The observed electron trapping thought to be associated with the isolated As$_{\rm Ga}$ must therefore come from other centers that may co-exist with As$_{\rm Ga}$ such as some O$_{\rm As}$-related defects. We find that GaAs can have multiple O-related defect centers, especially when prepared under As-rich conditions, which may include O$_{\rm As}$, O$_{i,{\rm brid}}$, O$_{i,{\rm tet}}$, O$_{\rm As}$-As$_{\rm Ga}$, and/or O$_{\rm As}$-2As$_{\rm Ga}$. Both O$_{\rm As}$ and O$_{\rm As}$-2As$_{\rm Ga}$ have a metastable and paramagnetic middle (neutral) charge state and exhibit certain characteristics of O-related defect centers observed in experiments; however, only O$_{\rm As}$-2As$_{\rm Ga}$ can be identified with the Ga--O--Ga or ``OX'' center, which confirms the previously proposed model for ``OX''. We have also explored possible radiative and non-radiative processes involving As$_{\rm Ga}$, Ga$_{\rm As}$, $V_{\rm Ga}$, O$_{\rm As}$, O$_{\rm As}$-As$_{\rm Ga}$, and O$_{\rm As}$-2As$_{\rm Ga}$ and found that the emission and absorption peak energies in defect-to-band transitions are all in the near infrared range. O$_{\rm As}$ and O$_{\rm As}$-2As$_{\rm Ga}$ have large nonradiative electron capture cross sections and can be effective carrier traps or recombination centers which has important implications for the performance of GaAs-based devices.

\begin{acknowledgments}

The author is grateful to Darshana Wickramaratne and Mark Turiansky for helpful discussions. This work used resources of the Center for Computationally Assisted Science and Technology (CCAST) at North Dakota State University, which was made possible in part by National Science Foundation Major Research Instrumentation (MRI) Award No.~2019077.

\end{acknowledgments}

\section*{Data Availability}

The data that support the findings of this study are included in this article and its appendix. Additional data are available from the author upon reasonable request.

\appendix

\section{Supporting figures and tables}\label{sec;app} 

This section includes defect local structures and formation energies and binding energies of defect complexes. 

\renewcommand\thetable{S1}
\begin{table*}
\caption{Stable charge states of defect complexes, their constituents, and binding energies ($E_{\rm b}$).}\label{tab;complex}
\begin{center}
\begin{ruledtabular}
\begin{tabular}{lclc}
Complex & Spin ($S$) & Constituents (Spin)$^\ast$ & $E_{\rm b}$ (eV)\\
\colrule
(Ga$_{\rm As}$-As$_{\rm Ga}$)$^0$& $0$ &Ga$_{\rm As}^{2-}$ ($S=0$) + As$_{\rm Ga}^{2+}$ ($S=0$)&2.61\\
(Ga$_{\rm As}$-As$_{\rm Ga}$)$^+$& $\frac{1}{2}$ &Ga$_{\rm As}^{-}$ ($S=\frac{1}{2}$) + As$_{\rm Ga}^{2+}$ ($S=0$)&1.99\\
(Ga$_{\rm As}$-As$_{\rm Ga}$)$^-$& $\frac{1}{2}$ &Ga$_{\rm As}^{2-}$ ($S=0$) + As$_{\rm Ga}^{+}$ ($S=\frac{1}{2}$)&1.74\\     
(Ga$_{\rm As}$-$V_{\rm As}$)$^0$& $\frac{1}{2}$ &Ga$_{\rm As}^{-}$ ($S=\frac{1}{2}$) + $V_{\rm As}^{+}$ ($S=0$)&0.74\\
(Ga$_{\rm As}$-$V_{\rm As}$)$^+$& $1$ & Ga$_{\rm As}^{0}$ ($S=1$) + $V_{\rm As}^{+}$ ($S=0$)&0.54\\
(Ga$_{\rm As}$-$V_{\rm As}$)$^-$& $0$ & Ga$_{\rm As}^{2-}$ ($S=0$) + $V_{\rm As}^{+}$ ($S=0$)&1.05\\
(As$_{\rm Ga}$-$V_{\rm Ga}$)$^0$& $\frac{1}{2}$ &As$_{\rm Ga}^{2+}$ ($S=0$) + $V_{\rm Ga}^{2-}$ ($S=\frac{1}{2}$)&1.40\\
(As$_{\rm Ga}$-$V_{\rm Ga}$)$^+$& $1$ &As$_{\rm Ga}^{2+}$ ($S=0$) + $V_{\rm Ga}^{-}$ ($S=1$)&0.98\\
(As$_{\rm Ga}$-$V_{\rm Ga}$)$^-$& $0$ &As$_{\rm Ga}^{2+}$ ($S=0$) + $V_{\rm Ga}^{3-}$ ($S=0$)&1.83\\
(As$_{\rm Ga}$-$V_{\rm Ga}$)$^{2-}$& $\frac{1}{2}$ &As$_{\rm Ga}^{+}$ ($S=\frac{1}{2}$) + $V_{\rm Ga}^{3-}$ ($S=0$)&1.03\\
($V_{\rm Ga}$-$V_{\rm As}$)$^0$& $1$ &$V_{\rm Ga}^{-}$ ($S=1$) + $V_{\rm As}^{+}$ ($S=0$)&1.46\\
($V_{\rm Ga}$-$V_{\rm As}$)$^-$& $\frac{1}{2}$ & $V_{\rm Ga}^{2-}$ ($S=\frac{1}{2}$) + $V_{\rm As}^{+}$ ($S=0$)&1.73\\
($V_{\rm Ga}$-$V_{\rm As}$)$^{2-}$& $0$ &$V_{\rm Ga}^{3-}$ ($S=0$) + $V_{\rm As}^{+}$ ($S=0$)&2.04\\
(O$_{\rm As}$-As$_{\rm Ga}$)$^0$& $\frac{1}{2}$ &O$_{\rm As}^{0}$ ($S=\frac{1}{2}$) + As$_{\rm Ga}^{0}$ ($S=0$)&1.65\\
(O$_{\rm As}$-As$_{\rm Ga}$)$^+$& $0$ &O$_{\rm As}^{+}$ ($S=0$) + As$_{\rm Ga}^{0}$ ($S=0$)&1.93 \\
(O$_{\rm As}$-As$_{\rm Ga}$)$^-$& $0$ &O$_{\rm As}^{-}$ ($S=0$) + As$_{\rm Ga}^{0}$ ($S=0$)&1.28\\
(O$_{\rm As}$-2As$_{\rm Ga}$)$^0$& $\frac{1}{2}$ &O$_{\rm As}^{0}$ ($S=\frac{1}{2}$) + As$_{\rm Ga}^{0}$ ($S=0$) + As$_{\rm Ga}^{0}$ ($S=0$)&2.83\\
(O$_{\rm As}$-2As$_{\rm Ga}$)$^+$& $0$ &O$_{\rm As}^{+}$ ($S=0$) + As$_{\rm Ga}^{0}$ ($S=0$) + As$_{\rm Ga}^{0}$ ($S=0$)&3.12\\
(O$_{\rm As}$-2As$_{\rm Ga}$)$^-$& $0$ & O$_{\rm As}^{-}$ ($S=0$) + As$_{\rm Ga}^{0}$ ($S=0$) + As$_{\rm Ga}^{0}$ ($S=0$)&2.79\\
\end{tabular}
\end{ruledtabular}
\end{center}

\begin{flushleft}
$^\ast$ Other isolated defects are Ga$_{\rm As}^+$ ($S=\frac{3}{2}$), $V_{\rm Ga}^0$ ($S=\frac{3}{2}$), $V_{\rm As}^-$ ($S=1$), $V_{\rm As}^{3-}$ ($S=0$), Ga$_i^+$ ($S=0$), Ga$_i^0$ ($S=\frac{1}{2}$), As$_i^+$ ($S=0$), As$_i^-$ ($S=0$), O$_{i,{\rm brid}}^0$ ($S=0$), O$_{i,{\rm tet}}^+$ ($S=\frac{1}{2}$), O$_{i,{\rm tet}}^0$ ($S=0$), O$_{i,{\rm tet}}^-$ ($S=\frac{1}{2}$), O$_{i,{\rm tet}}^{2-}$ ($S=0$), and (2O)$_{\rm As}^-$ ($S=0$).
\end{flushleft}

\end{table*}

\renewcommand{\thefigure}{S1}
\begin{figure*}[ht]
\centering
\includegraphics[width=\linewidth]{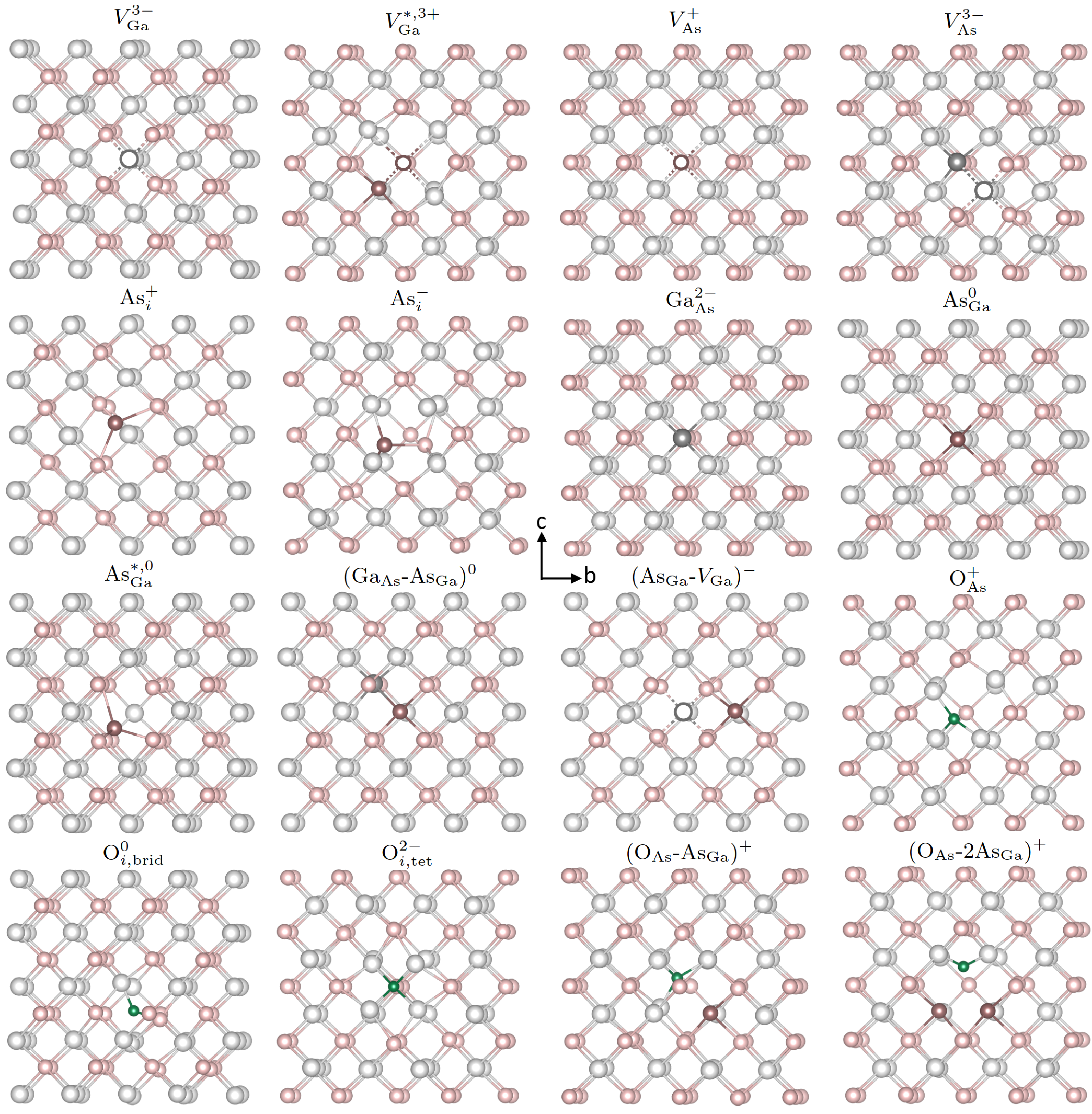}
\caption{Local structures of select native and oxygen-related defects in GaAs: $V_{\rm Ga}^{3-}$, $V_{\rm Ga}^{\ast,3+}$, $V_{\rm As}^{+}$, $V_{\rm As}^{3-}$, As$_i^+$, As$_i^-$, Ga$_{\rm As}^{2-}$, As$_{\rm Ga}^{0}$, As$_{\rm Ga}^{\ast,0}$, (Ga$_{\rm As}$-As$_{\rm Ga}$)$^0$ (i.e., a complex of Ga$_{\rm As}^{2-}$ and As$_{\rm Ga}^{2+}$), (As$_{\rm Ga}$-$V_{\rm Ga}$)$^-$ (i.e., a complex of As$_{\rm Ga}^+$ and $V_{\rm Ga}^{2-}$), O$_{\rm As}^+$, O$_{i,{\rm brid}}^0$, O$_{i,{\rm tet}}^{2-}$, (O$_{\rm As}$-As$_{\rm Ga}$)$^+$ (i.e., a complex of O$_{\rm As}^+$ and As$_{\rm Ga}^0$), and (O$_{\rm As}$-2As$_{\rm Ga}$)$^+$ (i.e., a complex of O$_{\rm As}^+$ and two As$_{\rm Ga}^0$). The large (gray) spheres are Ga and small (red) spheres are As where antisite Ga and As atoms are marked with darker colors for easy recognition; oxygen is represented by a green sphere. For defect complexes, see more information in Table~\ref{tab;complex}.}
\label{fig;struct}
\end{figure*}

\renewcommand{\thefigure}{S2}
\begin{figure}
\includegraphics*[width=0.9\linewidth]{figS2}
\caption{Migration barrier for the transition between the stable and metastable defect configurations.}
\label{fig;em} 
\end{figure}

\renewcommand{\thefigure}{S3}
\begin{figure*}
\includegraphics*[width=0.90\linewidth]{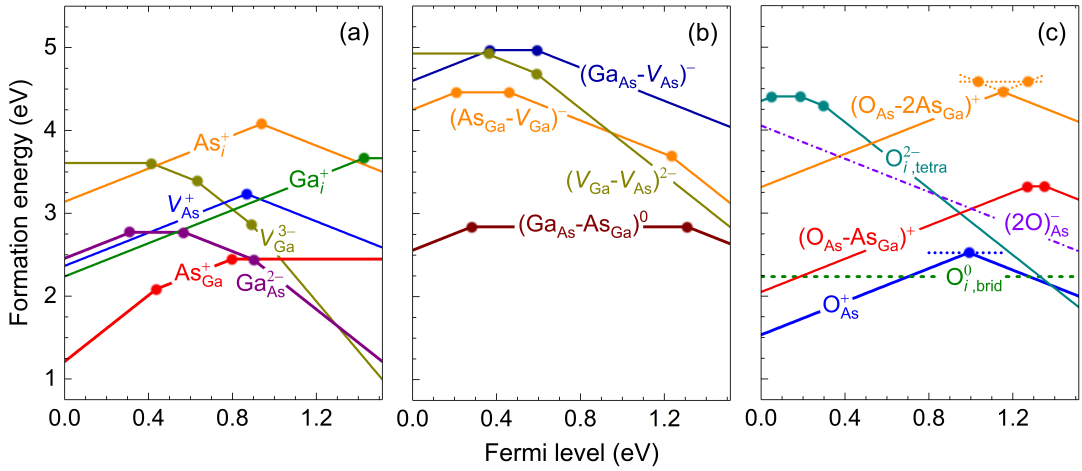}
\caption{Formation energies of (a) isolated native defects, (b) native defect complexes, and (c) O-related defects in GaAs, calculated at midpoint between the extreme Ga-rich and As-rich conditions. Large solid dots mark the defect levels.}
\label{fig;mid} 
\end{figure*}

\renewcommand{\thefigure}{S4}
\begin{figure}
\includegraphics*[width=0.9\linewidth]{figS4}
\caption{Emission rate of electrons from the ($+/0$) and ($0/-$) levels of O$_{\rm As}$-2As$_{\rm Ga}$ into the conduction band.}
\label{fig;emission} 
\end{figure}

%

\end{document}